\documentclass[11pt]{article}
\pdfoutput=1
\usepackage{geometry}
\usepackage{textcomp}
\usepackage{amsmath}
\begin{document}
\title{DEMoS Manifesto}
\author{Rasmus Ulslev Pedersen}
\maketitle
\begin{abstract}
This is a manifesto for DEMoS, which is a Distributed Embedded Modular System, but also a manifesto addressing the need for \textit{more} inter-/cross-disciplinary mastery of working knowledge related to \textit{installing} this class of systems in the real world. There is somehow room for yet another class of systems - complementary to existing embedded systems - complementing distributed operating systems - which takes on an interdisciplinary cyber-physical-materiality approach, a dedicated holistic perspective that recognizes the true value of interdisciplinary mastery vs. the implicit and overlooked expense of narrow intra-disciplinary focus dominating much of systems development (e.g. EE, CE, CS, SE, and IS). Interdisciplinary mastery yields its accumulated value across the development, deployment, use, re-use, and decommission phases for this class of systems: DEMoS is a system architected to be locally distributed, embedded, and modular as outlined herein and with the additional goals of human interdisciplinary mastery in this context: A potential set of goals for developing and applying DEMoS can be found in UN Resolution 70/1.
\end{abstract}
 In DEMoS \cite{Demos2016} (\textit{demos} also means \textit{'common people'}), the host SW tooling is mainly limited to multi-paradigm languages with open source implementations, e.g. the succinct F\# (in close competition with say C\#, but also other languages). For low-level HW description/verification, the Verilog HDL family is (with no serious competition) chosen: System Verilog emerging as the likely specific candidate. It is considered if transpiling (a sync./async. variant) of F\# to (System) Verilog is worth-while endeavor in one form or the other. On the embedded side, it would seem that any language with compiler support for the implemented instruction set is possible. However, multiple source language support (even from within the native .NET family) is most likely not practical on the target side (polyglot programming is harder in the embedded space!). In this case, F\# \cite{Syme2011} seems to offer a realistic compromise between the different programming paradigms (supporting both functional and/or OO programming.
 For PCB support of being distributed and modular, a system based on high-density interconnects is considered. Interconnects are currently thought to be based on Z-Ray\textsuperscript{\textregistered} 1.0 mm micro arrays, and I call it ZMOD. Considering also soldering, it, ZMOD, would make a kind of two-sided PCB available, while only demanding single sided SMD processes; not an irrelevant consideration from a maker perspective \cite{Hatch2014}. IO and some aspects of system control is thought to include use of PC/mobile phone audio ports with supporting target circuitry. For example, it opens up for programs or apps that support the the developer in some situations and the users in other scenarios. Just consider trouble-shooting complete (perhaps even remote) systems like drones, autonomous cars, robots, vessels or smaller power plant set-ups. Furthermore, is it considered what USB/JTAG solution to use. FTDI offers non-programmable and programmable USB solutions. Programmable USB solutions (e.g. VNC2) offer one way to deal with some IO mess (like audIO/UART/SPI/modems/WiFi/WiFi beacon-based protocols/BT/NFC/JTAG etc.), but with the annoyance of another SDK to learn. For low-speed/high-speed, small/wide, and wired/wireless interconnects to other cores, systems, transceivers, etc., a language independent specification is considered to be useful. For instance, to define mods based on the Z-Ray\textsuperscript{\textregistered} micro arrays, XML or JSON-LD seem good formats for this task. Another interconnect type that might prove useful is the well-known ethernet cable, as it is easy to cut into any desired length and connect to other "distant" (centimeters to meters away) DEMoS systems. Within one DEMoS system, a specialized processor, targeting the CIL instructions, called CILOP, is situated between DEMoS and the physical cores. One CILOP is similar to one Java Optimized Processor \cite{Schoeberl2010}. The (physical) cores, currently referred to as Thunder-cores (TCORE), are sought to have a non-rigid relationship to the operating system modules thus achieving distribution primarily within the \textit{embedded context}. One DEMoS could encompass many CILOPS, and those CILOPs could be run over multiple TCOREs, and those TCOREs could be spread across multiple FPGAs. It should be noted that a future, but both difficult and expensive, yet very interesting, option would be to "harden" the CILOP using services such as "The MOSIS Service". Coming back to the context, and thus the meaning of being distributed, is different for embedded systems than for \textit{general distributed (operating) systems} \cite{Jul1987}. Distributed execution - i.e. as in one robot - demands extreme care and understanding of the development environments/tools/processes as it is not trivial to develop multiple systems (with separate debug cables etc.) as one conceptional DEMoS. A basic un-distributed canonical implementation could consist of just one DEMoS running in one Thunder-core. Thunder-cores must support efficient average execution as well as real time applications. Needed ML/AI precision should be addressed by sensible FPU \cite{IEEE2008} support carefully balancing both FPGA cost and library implementation. Thunder-cores execute on an ISA currently aligned to ECMA-335, Partition III, because it is important to retain a close link to the fundamental binary execution units (.DLL and .exe). Flexible hardware is chosen in the form of affordable FPGAs/CPLDs, which allows more room for form to follow function, and function to follow form, in a wide spectrum of imaginable use cases with the developer possibly taking on an additional action research role. To support and exploit the marvelous FPGAs, specialized hardware is considered, such as optimized memory modules for garbage collection, stack fill/spill, context switching, process/app/OS migration, virtualization, and (large) (shared) caches; the latter most likely an important component to achieve meaningful distribution of data and computation. Considering the many pros and cons of the current market for affordable FPGAs/CPLDs (currently e.g. XILINX's Artix-7 family and Intel Programmable Solutions Group's MAX 10 series), the MAX 10 series seems the better way to proceed. 1 mm BGA pitch is anyway the right package to proceed with for both the MAX 10 as well as for the identified ISSI memory modules. In addition, at least currently, Quartus Prime seems to offer faster development flows enhancing the development experience (i.e. less time consuming).

 For realization in actual prototypes and products, additive manufacturing principles is a path worth serious pursuit: New software subscription models (e.g. from Autodesk) makes integrated 3D visualization/simulation of both electrical circuits and (cyber) physical product parts (i.e. here understood as a limited example of cyber-physical-materiality (cyberphysicalmateriality.org)) possible. In addition, DIN enclosures and DIN rails are good ways to encapsulate sets of (reusable) interconnected PCB modules. It has been considered to use DIP SPI ICs to make systematic testing and even support (e.g. firmware upgrades, some black-box functionality, or data-logging) manageable; even when deployed. It seems that Eurocircuits is a good PCB provider with international production, and initial testing (using EAGLE) came out satisfactory. Visual debugging/inspection at (an affordable, yet useful) 4-bit-nibble-level can be supported by just 4-5 leds together with some cobber trace contacts that can be triggered using say a small metal screwdriver. It is considered to focus on development and use of an AT-like bus command set perhaps with words (both as hex. 0xAC and ASCII "AC" as needed) like "AC"/"DC", "ADC"/"DAC", "ACE"/"ACED", or "EDDA" together with small operands to achieve some support for bus debugging, easy MSO triggering, useful readable tracing/logging, and data communication. A project hosting environment is thought to be at some \(\Lambda+\) gitlab/github repository. To remain flexible and bazaar-like, research encouraging measures are taken; i.e. using multi-letter (airport) codes (e.g. https://www.world-airport-codes.com/) to provide a wide layer of evolutionary development possible on top of DEMoS modules (SW/HW etc.) especially valuable to individual researchers, but also smaller R\&D teams. It is meant to strike a delicate balance between knowledge spill-overs (e.g. ecosystem thinking), yet allowing amble space for working on (time-consuming) "reference-able" research projects.
 I have outlined an architecture for a system:\\
 $DEMoS \Leftrightarrow CILOP \Leftrightarrow TCORE \Leftrightarrow FPGA \Leftrightarrow ZMOD$, which seems complicated at first, but the goal of interdisciplinary mastery (from cyber-physical systems \cite{Lee2008} to Micro Information Systems \cite{Pedersen2010}) is made possible by a setup which requires not many languages to master: mostly namely F\# and System Verilog (or similar substitutes on the SW and HW layers). It is possible that DEMoS can yield both enlightening research within a new branch of SE and also yield results in i.e. the Industrial Internet or areas such as many of those outlined in UN General Assembly resolution 70/1 (Transforming our world: the 2030 Agenda for Sustainable Development \cite{UN2015}). DEMoS \textit{is} cyber-physical-materiality and there is no \textit{edge} (as in "the edge of the Internet"; e.g. IoT) anymore.

\bibliographystyle{plain}
\bibliography{demos}
\end{document}